\documentclass[prl,aps,twocolumn,showpacs,preprintnumbers,amsmath,amssymb,superscriptaddress]{revtex4}
\usepackage{graphicx}
\begin{document}

\title{Spin-orbit coupling induced anisotropy effects in bimetallic
antiferromagnets: A route towards antiferromagnetic
spintronics}

\author{A. B. Shick}
\affiliation{Institute of Physics ASCR, v.v.i., Na Slovance 2, 182
21 Praha 8, Czech Republic}
\author{S. Khmelevskyi}
\affiliation{CMS, Vienna Univ. of Technology, Makartvilla,
Gusshausstr. 25, 1040 Vienna, Austria}
\author{O. N. Mryasov}
\affiliation{MINT, University of Alabama, Tuscaloosa, AL 35487, USA}
\author{J. Wunderlich}
\affiliation{Hitachi Cambridge Laboratory, Cambridge CB3 0HE, United
Kingdom}
\author{T. Jungwirth}
\affiliation{Institute of Physics ASCR, v.v.i., Cukrovarnick\'a 10,
162 53 Praha 6, Czech Republic} \affiliation{School of Physics and
Astronomy, University of Nottingham, Nottingham NG7 2RD, United
Kingdom}

\begin{abstract}
Magnetic anisotropy phenomena in bimetallic antiferromagnets
Mn$_2$Au and MnIr are studied by first-principles density functional
theory calculations. We find strong and lattice-parameter dependent
magnetic anisotropies of the ground state energy, chemical
potential, and density of states, and attribute these anisotropies
to combined effects of large moment on the Mn 3$d$ shell and large
spin-orbit coupling on the 5$d$ shell of the noble metal. Large
magnitudes of the proposed effects can open a route towards
spintronics in compensated antiferromagnets without involving
ferromagnetic elements.

\end{abstract}
\date{\today}
\maketitle


The introduction in 1991 of a hard-drive read-head based on the
ferromagnetic anisotropic magnetoresistance  (AMR) effect
represented a major step in the field which later on became known as
spintronics~\cite{Chappert:2007_a}. The basic physics of the AMR,
namely the bulk nature of this magneto-transport effect and its
subtle spin-orbit coupling (SOC) origin, have limited the magnitude
of the effect and the potential of scaling the AMR devices into
small dimensions required for current high-density magnetic storage
applications. These limitations were partly overcome by the
discovery of the giant and tunneling magnetoresistance   (GMR and
TMR) effects which are interface transport phenomena and which rely
primarily on the more robust magnetic exchange interaction.
To date, studies of magnetoresistive effects potentially suitable
for AFM spintronics have focused on AFM counterparts to the GMR
~\cite{Nunez:2006_a,Haney:2006_b,Duine:2007_a}. While the viability
of this approach is yet to be discerned it has been acknowledged
that the requirements on the structural quality and the coherence of
transport through interfaces in the GMR devices are significantly
more stringent in the case of AFMs.

This paper aims to open an alternative route towards AFM spintronics
which reintroduces the leading role of SOC. In this approach,
the stringent requirements on the GMR/TMR are circumvented by
considering instead the tunneling AMR (TAMR)  and the
Coulomb-blockade AMR (CBAMR)  effects. We also demonstrate that SOC
can be employed to control magnetic anisotropies in the AFM in a way
that leads to reorientation of the staggered moments required for
observing the above anisotropic transport effects.

Previous studies of the
TAMR~\cite{Gould:2004_a,Brey:2004_b,Shick:2006_a,Gao:2007_a,Park:2008_a}
and CBAMR~\cite{Wunderlich:2006_a,Bernand-Mantel:2009_a} in
ferromagnets have shown that anisotropic magnetoresistance phenomena
can be extended from bulk to nanoscale devices, can have large
magnitudes, and do not require spin-coherent transport throughout
the structure.
In transition metal ferromagnets a generic principle has been
outlined, based on studies of magneto-crystalline anisotropies and
of the TAMR and CBAMR
~\cite{Wunderlich:2006_a,Shick:2008_a,Park:2008_a}, that the
magnetic anisotropy phenomena are maximized in bimetallic systems
combining large spontaneous moments on the 3$d$ shell of a
transition metal and large magnetic susceptibility and SOC on the
5$d$-shell of a noble metal. Since Mn carries the largest  moment
among transition metals and most of the bimetallic alloys containing
Mn order antiferromagnetically, the goals of strong magnetic
anisotropy phenomena and of AFM spintronics  appear to merge
naturally together. In our relativistic {\em ab initio} study we
consider the Mn$_2$Au AFM for which recent theoretical calculations
predicted~\cite{Khmelevskyi:2008_a} record N\'eel temperature
($>1500$~K) among  Mn-based AFM alloys. The generic nature of the
proposed anisotropy phenomena is confirmed by calculations in the
conventional bimetallic AFM MnIr.

We begin with a detail analysis of magneto-crystalline anisotropies
in Mn$_2$Au. Calculations are performed  considering the
MoSi$_2$-type bct structure with experimental lattice parameters, $a
=6.291$~a.u. and $c=16.142$~a.u. The tetragonal unit cell with two
formula units (f.u.) and an AFM arrangement of magnetic moments are
shown in Fig.~1. This staggered magnetic structure was obtained from
the theoretical study~\cite{Khmelevskyi:2008_a} of the magnetic
interactions, and leads to the AFM stacking of the Mn layers with no
frustration.

\begin{figure}[htbp]
\includegraphics[angle=270,width=1.0\columnwidth,clip]{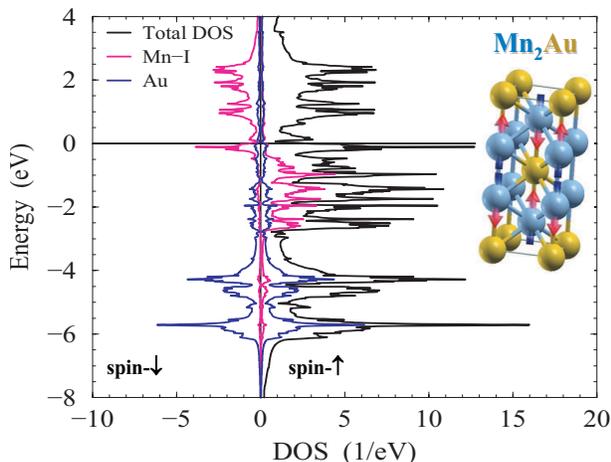}

\vspace*{-0.5cm} \caption{Total DOS (per f.u.), spin-resolved
Au-atom and Mn-atom projected DOSs, and a schematic view of the
crystal and AFM structure  of Mn$_2$Au. (Mn-I denotes one of two
Mn-sublattices with opposite spin-polarizations.)
} \label{dosfig}
\end{figure}

We employ the relativistic version of the full-potential linearized
augmented-planewave (FP-LAPW) method in the local spin-density
approximation. The SO coupling is included in a self-consistent
second-variational procedure~\cite{Shick:1997_a}. This approach is
typically very accurate for itinerant metallic systems. The
calculated total and atom resolved DOSs for moments aligned with the
[001] axis are shown in Fig.~1. We find that Mn atoms carry spin
$M_S=\pm3.2~\mu_B$ and orbital $M_L=\mp0.013~\mu_B$ magnetic moments
and there are no magnetic moments on the Au atoms. When staggered
magnetization is aligned along the [100]/[010]-axis there is no
change in $M_S$ of Mn and the magnitude of $|M_L|=0.007~\mu_B$ is
reduced.

\begin{table}[htbp]
\begin{center}
\begin{tabular}{c|c|c|c|c}
\hline &$K_{2\perp}$&$K_{4\perp}$&$K_{4\parallel}$&
$K^{\ast}_{2\parallel}$\\
\hline
Mn$_2$Au & -2.44 & 0.02 & 0.01 & 0.07  \\
Au   &-2.72  &0.01  &0.01  & 0.08       \\
Mn$_2$&0.28 &0.01  & 0.00 &-0.01  \\
\hline
\end{tabular}
\label{table1}
\end{center}
\caption{Relativistic full-potential DFT calculations of the
uniaxial $K_2$, and forth-order out-of-plane, and in-plane $K_{4
\perp},K_{4 ||}$ MAE constants (total and element specific in meV
per f.u.) of Mn$_2$Au. Additional uniaxial MAE
$K^{\ast}_{2\parallel}$ (in meV per f.u. per 1\% strain) is induced
by in-plane strain along the [100] crystal direction (see text).}
\end{table}

The magneto-crystalline anisotropy energy (MAE) is evaluated using
the torque method which is implemented in the FP-LAPW basis
~\cite{Shick:2008_a}. For the tetragonal symmetry case, the
phenomenological total energy dependence on the spin quantization
direction reads,
\begin{equation}
\label{torque} E(\theta,\phi) = K_{2\perp} \sin^2\theta
+ K_{4 \perp} \sin^4\theta + K_{4\parallel} \sin^4\theta \cos4\phi
\end{equation}
and the corresponding torque is given by, $T(\theta,\phi) =
dE(\theta,\phi)/d\theta = \sin2\theta \; [K_{2\perp}  + 2 (K_{4
\perp} + K_{4\parallel} \cos4\phi)  \sin^2\theta] $. Here
$K_{2\perp}$ is the uniaxial MAE constant, and $K_{4 \perp}$ and
$K_{4\parallel}$ are the forth-order out-of-plane and in-plane MAE
constants, respectively. The advantage of this approach is that it
allows us to split the total MAE into the element-specific
contributions from different atoms in the unit cell. The total and element specific anisotropy constants are shown in
Tab.~I. They are calculated with accuracy better than 0.01 meV/f.u.

The largest uniaxial anisotropy constant, $K_{2\perp}$, is strong
and clearly dominated by contributions from the Au sublattice. The
leading role in the MAE of Au-atoms which carry zero net moment may
appear counterintuitive and requires more detail inspection. In the
AFM state the Au orbitals point towards the neighboring Mn moments
with opposite orientations and strong hybridization between Mn and
Au creates oppositely polarized parts of the Au orbitals.
Considering separately exchange fields produced by the two Mn
sublattices, $\alpha=\uparrow,\downarrow$, with opposite strongly
localized moments  (${\bf \pm M^{3d}}$), each of the sublattices
induces magnetic moment ${\bf {M}}^{5d}_{\alpha}$ on the itinerant
5$d$ electrons of Au atoms. We can
write~\cite{Mryasov:2005_a,Shick:2008_a}
\begin{eqnarray}
\label{5dm} {\bf M}^{5d}_{\alpha}=\chi \sum_{i} J_{3d-5d}^{i,\alpha}
{\bf M}^{3d}_{\alpha,i} \, ,
\end{eqnarray}
where we sum over the Mn atoms in the sublattice $\alpha$,
$J_{3d-5d}^{i,\alpha}$ is the exchange interaction between the
$i-{\rm th}$ Mn atom from the sublattice $\alpha$ and the Au atom,
and $\chi$ is the local spin susceptibility of the Au atom. Strong
SOC on Au yields the MAE contributions due to ${\bf
{M}}^{5d}_{\uparrow,\downarrow}$. For the uniaxial term it can be
written as,
$E_{A,\alpha}^{5d}=-k_{2\perp}^{5d}({M}_{\alpha}^{5d,z})^2$, where
the renormalized anisotropy constant $k_{2\perp}^{5d}$ is
proportional to the square of the SOC parameter $\xi_{5d}^2$. Using
Eq.~(\ref{5dm}) we can write,
\begin{eqnarray}
\label{5dA} E_{A, \alpha}^{5d} = -k_{2}^{5d} \chi^2 \sum_{ij}
{J}_{3d-5d}^{i,\alpha} {J}_{3d-5d}^{j,\alpha} {M}^{3d,z}_{\alpha,i}
{M}^{3d,z}_{\alpha,j} \, .
\end{eqnarray}
Summing Eq.~(\ref{5dm}) over the Mn sublattices $\alpha$, we get
zero magnetic moment which complies with the overall time-reversal
symmetry of the AFM band structure. The total MAE due to Au,
$\sum_{\alpha=\uparrow,\downarrow} E_{A, \alpha}^{5d}$, is non-zero,
however, as seen from Eq.~(\ref{5dA}). It originates from a
combination of strong 5$d$ SOC ($k_{2\perp}^{5d}$), strong exchange
splitting induced by the 3$d$ magnetic element (${J}_{3d-5d} {\bf
M}^{3d}$), and the enhancement of the local spin susceptibility
$\chi$. Eq.~(\ref{5dA}) implies that MAE is
predominantly governed by two-site anisotropy and is proportional to
the number of bonds between the Au and Mn atoms.

Large magneto-crystalline anisotropies in bimetallic AFMs have
already been reported in {\em ab initio} studies of the common AFM
MnIr. The key role of polarized orbitals of the noble metal has not,
however, been identified in these works. To test that the physics
described in the previous paragraph is generic, we have examined the
element specific MAE of MnIr. In the calculations we considered the
experimental lattice constants of the L1$_0$ MnIr and the unit cell
with two f.u.'s in a checkerboard collinear AFM
structure~\cite{Umetsu:2006_a,Szunyogh:2008_a}.
For the total MAE
($K_{2\perp}+K_{ 4 \perp} + K_{4\parallel}$) we obtained
-3.365~meV/f.u. (-6.73~meV per unit cell). This number is in  good
agreement with previous results (-7.05 meV per unit cell in
Ref.~\onlinecite{Umetsu:2006_a}, -6.81 meV per unit cell in
Ref.~\onlinecite{Szunyogh:2008_a}). Our element-specific decomposition of
the MAE shows that, similarly to the case of Mn$_2$Au, the MAE is
dominated by the contribution from the noble metal which again
carries zero net moment in MnIr. For the leading  uniaxial
MAE term $K_{2\perp}$ we obtained a contribution of -3.38~meV/f.u.
from Ir and a much weaker contribution of  0.024 meV/f.u from Mn.

We now turn to the discussion of the magneto-transport effects in AFMs. The AMR effects are even in magnetic moment and, similar
to the MAE, can be finite  in the compensated AFM.
In what follows we provide estimates of the AMR in the Coulomb
blockade and tunneling regimes. To date TAMR and CBAMR have been
observed only in ferromagnets. We argue that both these effects are
present and can be large in compensated AFMs.

The CBAMR builds  on a general principle that transport through an
electronic device depends on positions of electro-chemical
potentials in the relevant electrodes. Transport and  band structure
with SOC are in this case directly linked via the anisotropy of the
chemical potential with respect to the orientation of magnetic
moments~\cite{Wunderlich:2006_a,Tran:2009_a}. Single electron
transistors (SETs) are arguably the most sensitive devices to detect
this effect.
Large magneto-Coulomb oscillations of the conductance are induced by
rotating magnetic moments, whenever the changes in the chemical
potential become comparable to the single-electron charging energy
of the central island in the SET~\cite{Wunderlich:2006_a}. Our
calculated difference between chemical potentials for the staggered
magnetization aligned along the [110] axis and the [001] axis is
-2.5~meV for Mn$_2$Au and 3.2~meV for MnIr.  This is comparable to
chemical potential anisotropies in ferromagnetic semiconductor
(Ga,Mn)As and in bimetallic ferromagnets~\cite{Wunderlich:2006_a}.
The AFM CBAMR or a magnetoresistance of other AFM devices which are
sensitive to changes of  electro-chemical potentials of the order of
a few mV's should therefore be readily detectable.

For FM tunneling devices it has been demonstrated that a useful
qualitative analysis of the differential TAMR can be obtained by
considering its proportionality to the energy dependent anisotropy
in the density of states  (ADOS) in the magnetic electrode with
respect to the crystallographic orientation of magnetic
moments~\cite{Gould:2004_a,Shick:2006_a}. Anisotropy in the
group-velocity weighted tunneling density of states (ATDOS) is
considered instead of the ADOS when referring to tunnel junctions
with high quality interfaces, i.e., high degree of in-plane momentum
conservation during tunneling~\cite{Gould:2004_a,Shick:2006_a} We
emphasize that in either case these calculations are only
approximate as they  neglect the anisotropy of tunneling matrix
elements between wavefunctions in the magnetic and nonmagnetic
electrodes of the TAMR structure~\cite{Moser:2006_a}.

Calculated energy-dependent ADOS and ATDOS for  Mn$_2$Au and
staggered moments aligned with the easy [110] and hard [001]
staggered magnetization axes are plotted in Fig.~2(a).  The ADOS can
be as high as 50\% in the vicinity of the Fermi level and oscillates
strongly with energy. This implies a sizable TAMR whose magnitude
and sign are bias dependent, similar to the experimentally observed
TAMR characteristics in bimetallic
ferromagnets~\cite{Gao:2007_a,Park:2008_a}. The calculated ATDOS is
also large and shows weaker dependence on energy around the Fermi
level. In MnIr the ADOS between the easy [100] and hard [001]
staggered magnetization axes is weaker but still reaching 10\%, as
shown in Fig.~2(a). Note that for staggered moment reorientations
between the easy and hard axes, MnIr is expected to display weaker
TAMR of the two considered AFMs while the corresponding magnetic
anisotropy constant $K_{2\perp}$ and the anisotropy in the chemical
potential are stronger in MnIr. This illustrates that despite their
common SOC origin the different magnetic anisotropy phenomena can
behave to some extent independently.

\begin{figure}[htbp]
\includegraphics[angle=0,width=1.0\columnwidth,clip]{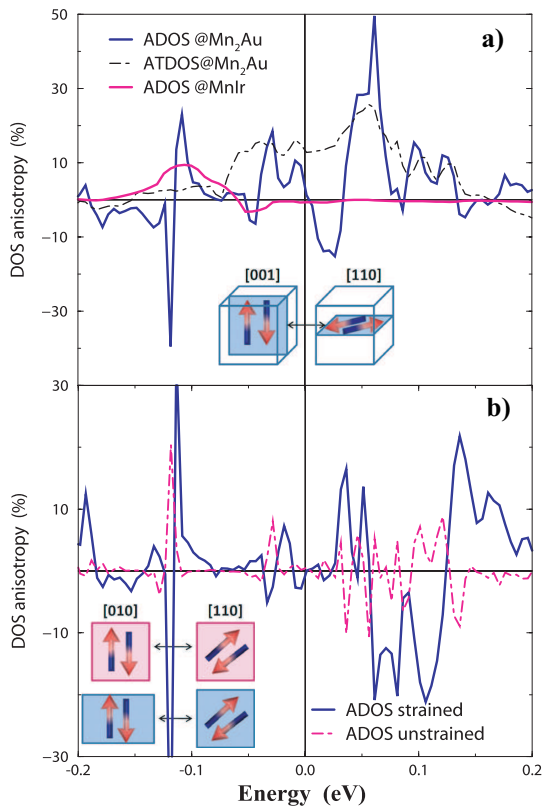}
\vspace*{-0.5cm} \caption{(a) DOS anisotropies for
the hard ([001]) and easy ([110]) axes of Mn$_2$Au,
and the hard ([001]) and easy ([100]) axes of MnIr.
(b) In-plane DOS anisotropies for staggered moment aligned along the
easy axis ([010]) of the strained Mn$_2$Au crystal and along the
easy axis ([110]) of the unstrained Mn$_2$Au.
The
energy is measured from the Fermi level.
} \label{dosfig2}
\end{figure}

The absence of a net magnetization makes AFMs rigid to magnetic
fields and therefore one needs to resort to manipulation of the
staggered moments via internal fields. One approach, previously
reported in the literature, is based on exchange-spring
effects~\cite{Scholl:2004_a} occurring when the AFM is interfaced
with a ferromagnet. This technique can be used to trigger the
reorientation between the easy [110]-axis and hard [001]-axes which
produces the TAMR effects predicted in Fig.~2(a). Instead of this
exchange field method we now return to the magneto-crystalline
anisotropy and demonstrate that the staggered moments can be
controlled via this internal field without involving ferromagnets.
Specifically we consider the dependence of the MAE and of the
corresponding in-plane easy-axis orientation in the AFM on the
lattice strain.  (This can be controlled externally by, e.g., a
piezoelectric stressor~\cite{Rushforth:2008_a}.)

As shown in Tab.~I, the in-plane easy-axis of unstrained Mn$_2$Au  lies along the [110]
(or [1$\bar{1}$0]) crystal direction. By applying a
sufficiently strong strain along one of the cube edges, the
easy-axis, and therefore also the staggered moments, will rotate towards the [100] or [010] direction, depending on the sign of the strain. To model
this effect we elongated/contracted  the unit cell along the
[100]/[010]-axis, keeping the unit cell volume fixed. The
corresponding crystal structure becomes orthorombic and the second
order uniaxial anisotropy terms are given by
$\sin^2\theta[K_{2\perp}+ K^{\ast}_{2\parallel}\cos2\phi]$.

Assuming strains up to 1\% we found negligible changes in the
out-of-plane uniaxial constant $K_{2\perp}$ and in the fourth order
constants $K_{4\perp}$ and $K_{4\parallel}$.  For the strain-induced
in-plane uniaxial MAE we obtained
$K^{\ast}_{2\parallel}=0.07$~meV/f.u. per 1\% strain and again the
dominant contribution comes from the Au-atom (0.08 meV/f.u. per
1\%). The easy-axis shifts from the in-plane diagonal to the [010]
direction when $K_{2\parallel}>2K_{4\parallel}$. It means that only
a fraction of a per cent strain is required  to rotate the staggered
moments between the [110] and [010] axes. These crystallographic
directions are non-equivalent in the tetragonal  Mn$_2$Au and can
yield a sizable TAMR, as shown by the corresponding in-plane ADOS in
Fig.~2(b). Note that the detailed TAMR characteristics can be
modified by the strain as illustrated by the other ADOS curve in
Fig.2(b) which corresponds to the relative difference between the
DOS in the unstrained case and moments along the [110] axis, and the
DOS in 1\% strained case and moments along the [010] axis.
Nevertheless, the primary role of the strain is in the MAE where it
triggers the reorientation of the magnetic easy-axis.


In conclusion, we have proposed a set of relativistic magnetoresistance effects which can open
a route towards AFM spintronics and which can be realized in AFMs alone without involving
ferromagnetic elements.
Our previous  works in which TAMR and CBAMR were predicted for
transition metal ferromagnets~\cite{Shick:2006_a,Wunderlich:2006_a}
using similar relativistic {\em ab initio} techniques have been
subsequently confirmed by
experiments~\cite{Gao:2007_a,Park:2008_a,Bernand-Mantel:2009_a}.
This together with the identified microscopic physics behind the
magnetic anisotropy effects in compensated AFMs gives us confidence
that predictions presented in this paper are realistic.

We acknowledge stimulating discussions with Bryan Gallagher, Allan
MacDonald, and Jan Ma\v{s}ek and financial support from EU Grants
FP7-215368 SemiSpinNet, FP7-214499 NAMASTE, Austrian Grant MOEL
OEFG, Czech Republic Grants GACR ON/06/E001, FON/06/E002, GACR
202/07/0456, GACR P204/10/0330, GAAV IAA100100912, AV0Z10100520,
AV0Z10100521, KAN400100652, LC510, and Preamium Academiae.

\vspace*{-0.5cm}

\end{document}